\documentclass[prl,aps,twocolumn,citeautoscript,reprint,superscriptaddress]{revtex4-2}
\usepackage{graphicx}
\usepackage{bm,amsmath,amssymb,wasysym,amsfonts,dcolumn}
\usepackage[colorlinks=true, urlcolor=blue, linkcolor=blue, citecolor=blue, pdfborder={0 0 0}]{hyperref}
\usepackage[usenames,dvipsnames]{xcolor}
\usepackage{mathrsfs}
\usepackage{multirow}
\usepackage{makecell}

\newcommand{\DOI}[1]{\href{https://doi.org/#1}}

\begin{document}

\title{Ferromagnetic interface engineering of spin-charge conversion in RuO$_2$}

\author{Dongchao Yang}
\altaffiliation{These authors contributed equally to this study.}
\affiliation{School of Physics Science and Engineering, Tongji University, Shanghai 200092, China}

\author{Zhaoqing Li}
\altaffiliation{These authors contributed equally to this study.}
\affiliation{Interdisciplinary Center for Theoretical Physics and Information Sciences, Fudan University, Shanghai 200433, China}

\author{Yu Dai}
\affiliation{Anhui Provincial Key Laboratory of Magnetic Functional Materials and Devices, School of Materials Science and Engineering, Anhui University, Hefei 230601, China}

\author{Lili Lang}
\affiliation{Shanghai Institute of Microsystem and Information Technology, Chinese Academy of Sciences, 865 Changning Road, Shanghai 200050 China}

\author{Zhong Shi}
\email{shizhong@tongji.edu.cn}
\affiliation{School of Physics Science and Engineering, Tongji University, Shanghai 200092, China}

\author{Zhe Yuan}
\email{yuanz@fudan.edu.cn}
\affiliation{Interdisciplinary Center for Theoretical Physics and Information Sciences, Fudan University, Shanghai 200433, China}

\author{Shi-Ming Zhou}
\affiliation{School of Physics Science and Engineering, Tongji University, Shanghai 200092, China}

\date{\today}
	
\begin{abstract}
Spin-orbit torque efficiency is conventionally fixed by bulk materials. $D$-wave altermagnets introduce an additional nonrelativistic spin-charge conversion channel beyond inverse spin-Hall effect. Using prototypical candidate RuO$_2$ as an example, we show that the adjacent ferromagnet alone can dictate both the magnitude and sign of spin-charge conversion. Spin-pumping measurements on RuO$_2$/Y$_3$Fe$_5$O$_{12}$ (YIG) and RuO$_2$/Ni$_{80}$Fe$_{20}$ (Py) bilayers yield opposite effective spin-Hall angles that persist across crystalline and polycrystalline RuO$_2$. Inserting an ultrathin Au spacer at the RuO$_2$/YIG interface reverses the signal, envidencing a dominant interfacial inverse Rashba-Edelstein effect, whereas RuO$_2$/Py is governed by bulk inverse spin-Hall effect. First-principles calculations trace this dichotomy to interface-selective band hybridization: Rashba surface states survive at the insulating YIG contact yet are quenched by metallic Py. Our findings establish ferromagnetic interfacing as a deterministic knob for tailoring spin-charge conversion in altermagnetic oxides, paving the way to field-free, low-dissipation spintronic memory devices.
\end{abstract}
	
\maketitle

{\color{red}\it Introduction.---}Spin-orbit torque (SOT)~\cite{manchon2019current,song2021spin} offers an energy-efficient route to magnetization switching in spintronic devices, typically mediated by the spin Hall effect (SHE)~\cite{liu2012spin,liu2012current} in heavy metals. The SHE, however, is bound by symmetry: spin current, spin polarization, and charge current must be mutually orthogonal~\cite{sinova2015spin}. In perpendicular magnetized bilayers, this condition prevents deterministic switching without an auxiliary in-plane field. Symmetry can be broken by sophisticated fabrication techniques, such as wedge-shaped layers \cite{cai2017electric} or spatially inhomogeneous ferromagnetic layers~\cite{tang2020bulk}. But a simpler strategy is to inject an out-of-plane ($z$-polarized) spin current. Such collinear spin currents can be produced in low-symmetry two-dimensional crystals~\cite{macneill2017control}, ferromagnetic alloy \cite{yang2024highly} or via the anisotropic spin splitting effect (ASSE)~\cite{naka2019spin,yuan2020giant,hayami2020bottom,yuan2021strong,naka2021perovskite,gonzlez2021efficient} recently uncovered in altermagnets~\cite{smejkal2022beyond,smejkal2022emerging,savitsky2024researchers}. Owing to its large predicted ASSE and simple rutile structure, RuO$_2$ is a leading candidate in this class.
	
Charge-to-spin conversion driven by the ASSE has been demonstrated in RuO$_2$ by spin-torque ferromagnetic resonance (FMR) and harmonic Hall measurements~\cite{bai2022observation,karube2022observation,bose2022tilted,zhang2024simultaneous,guo2024direct}, while its inverse process has been detected by spin Seebeck effect~\cite{bai2023efficient,liao2024separation}, terahertz emission~\cite{liu2023inverse}, and spin-pumping experiments~\cite{guo2024direct,wang2024inverse}. Direct measurements of the ASSE-induced torque have been reported~\cite{bai2022observation,karube2022observation}, and tuned by rotating the Néel vector with field annealing. Competing contributions from the SHE and its inverse have also been observed~\cite{liao2024separation,wang2024inverse,sun2017dirac}, but the shared angular dependence of ASSE- and SHE-generated spin currents makes their separation difficult~\cite{bose2022tilted,wang2024inverse}. While reciprocal measurements on RuO$_2$/Ni$_{80}$Fe$_{20}$ (Py) bilayers assigned a positive spin Hall angle (SHA)~\cite{wang2024inverse}, independent studies of RuO$_2$/Y$_3$Fe$_5$O$_{12}$ (YIG) systems reported a negative SHA~\cite{wang2025robust}. This puzzling sign discrepancy further complicates the ongoing debate over altermagnetism in RuO$_2$~\cite{fedchenko2024observation,smolyanyuk2024fragility,zhu2019anomalous,berlijn2017itinerant,feng2022an,hiraishi2024nonmagnetic,kessler2024absence,wenzel2025fermi,jovic2018dirac,zhang2025probing,kiefer2025crystal,liu2024absence,jeong2024altermagnetic,gregory2022strain,qian2025fragile}. Disentangling possible additional contributions is therefore indispensable for unambiguous experimental verification of the ASSE and for clarifying the magnetic ground state of RuO$_2$.

\begin{figure*}[t]
\centering
\includegraphics[width=14cm]{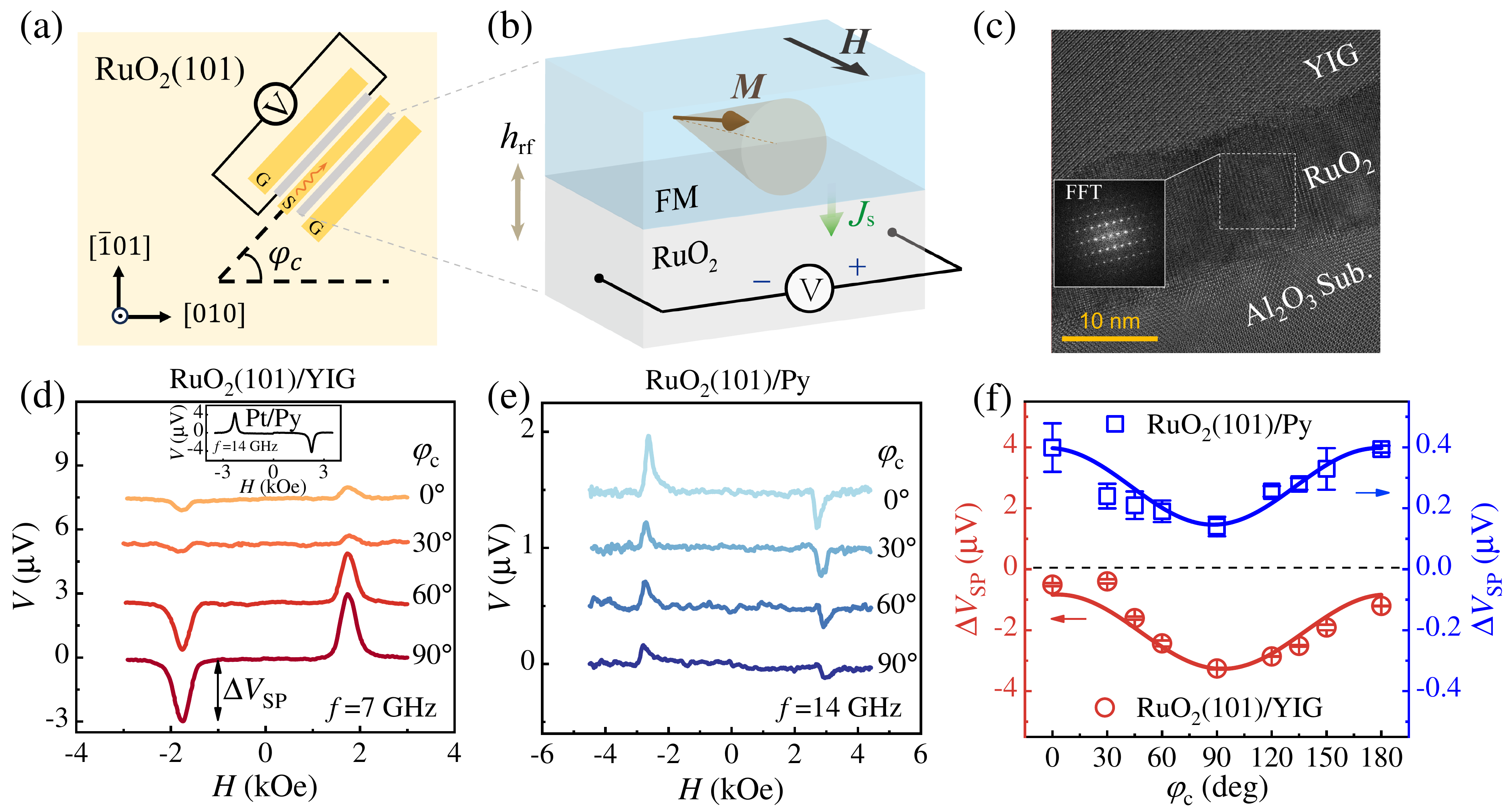}
\caption{(a) Schematic of the RuO$_2$(101)/FM bilayer. The gray bar indicates the etched stripe for spin-pumping detection, positioned between a coplanar waveguide. (b) Experimental arrangement. Microwave excites FMR in the FM layer and spin-charge conversion is read out as a dc voltage across the stripe. (c) Cross-sectional TEM image of the RuO$_2$(101)/YIG bilayer. Inset: FFT of a selected RuO$_2$ region confirming single-crystal quality. (d) Measured voltage of the RuO$_2$(101)(15 nm)/YIG(42 nm) bilayers as a function of applied magnetic field. The arrow marks the peak value $\Delta V_{\rm SP}$. Inset: reference signal from a polycrystalline Pt(4~nm)/Py(6~nm) bilayer recorded under the identical conditions. (e) Corresponding data for the RuO$_2$(101)(15 nm)/Py(6 nm) bilayers; note the sign reversal with respect to (d). (f) Angular dependence of $\Delta V_{\rm SP}$ for both systems. Symbols are the experimental data and the solid lines illustrate $\textrm{cos}^{2}\varphi_c$ fits. The error bars are defined as $(\vert\Delta V_{\rm SP} (H_+ )\vert-\vert\Delta V_{\rm SP} (H_- )\vert)/2$ with $\Delta V_{\rm SP} (H_+)$ and $\Delta V_{\rm SP} (H_-)$ measured at the positive and negative FMR fields, respectively.
\label{fig1} }
\end{figure*}
	
Here we resolve the conflicting SHAs by identifying an inverse Rashba-Edelstein effect (IREE) that dominates spin-charge conversion at the RuO$_2$/YIG interface. Spin-pumping measurements reveal a robust, negative effective SHA in RuO$_2$/YIG, which is reversed when a thin Au interlayer is inserted to suppress this interfacial effect. RuO$_2$/Py displays a positive SHA governed by bulk inverse SHE (ISHE). First-principles calculations show that Rashba-type surface states, generated once spin-orbit coupling (SOC) removes the topological protection, survive at the RuO$_2$/YIG interface owing to negligible band hybridization, but are quenched at the RuO$_2$/Py interface. These findings establish interface engineering as a powerful knob for tailoring spin-orbit phenomena in altermagnetic systems and chart a route to field-free, energy-efficient magnetization control. 

{\color{red}\it Spin-charge conversion in RuO$_2$/FM bilayers.---}We prepared RuO$_2$(101)/YIG and RuO$_2$(101)/Py bilayers to quantify spin-charge conversion via spin pumping. First, 15-nm-thick (101)-oriented RuO$_2$ films were epitaxially grown on Al$_2$O$_3$($1\bar{1}02$) substrates by pulsed-laser deposition at 450 $^\circ\mathrm{C}$. A 42-nm-thick polycrystalline YIG layer was then deposited at 350 $^\circ\mathrm{C}$ and post-annealed in air at 850 $^\circ\mathrm{C}$ for 8 minutes. High-temperature annealing preserves the RuO$_2$ crystallinity~\cite{SM}. For the RuO$_2$(101)/Py stack, a 6-nm-thick polycrystalline Py layer was sputtered at room temperature. Both heterostructures were lithographically patterned into $20~\mu{\rm m}\times 2~{\rm mm}$ stripes along selected RuO$_2$ crystallographic directions, as sketched in Fig.~\ref{fig1}(a). Here $\varphi_c$ denotes the angle between the stripe axis and the [010] direction. High-resolution transmission electron microscopy (TEM) image in Fig.~\ref{fig1}(c) confirms the single-crystal quality of RuO$_2$(101); further fabrication details are provided in the Supplemental Material~\cite{SM}.

Spin pumping was carried out in the configuration sketched in Fig.~\ref{fig1}(b). A microwave magnetic field introduced by a coplanar waveguide (50~$\mu$m from the sample) drives FMR in the ferromagnet (FM) layer, and the resulting spin current is converted into a dc voltage across the RuO$_2$ stripe. Throughout the measurement, an external magnetic field is applied perpendicular to the stripe, irrespective of $\varphi_c$. Figure~\ref{fig1}(d) shows the voltage versus the applied magnetic field $H$ for the RuO$_2$/YIG bilayer. Sharp peaks at $H=\pm1.8$~kOe mark FMR; the sign reversal at opposite fields reflects the inverted spin-current polarization. Control measurements on single Py layers confirm negligible self-induced effects~\cite{SM}. The largest and smallest signals occur at $\varphi_c=90~^\circ$ and $\varphi_c=0~^\circ$, respectively.

\begin{figure*}[htbp]
\centering
\includegraphics[width=14cm]{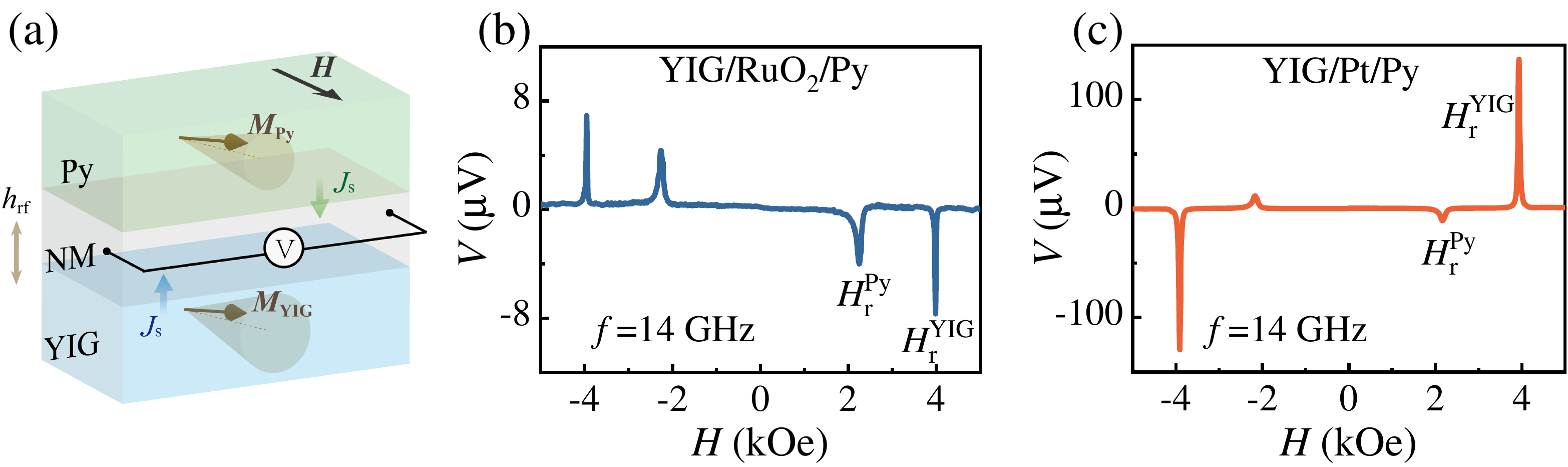}
\caption{(a) Device sketch. Distinct FMR fields of YIG and Py allow spin current to be injected into the nonmagnetic layer from either side within one measurement. (b) Spin-pumping signals for YIG(42 nm)/RuO$_2$(8 nm)/Py(10 nm) trilayer as a function of applied magnetic field. (c) The same measurement for the reference sample YIG(42 nm)/Pt(6 nm)/Py(10 nm). \label{fig2}}
\end{figure*}
	
The RuO$_2$/Py bilayer exhibits FMR at $\pm2.8$~kOe, but the voltage polarity is reversed relative to RuO$_2$/YIG [Fig.~\ref{fig1}(e)]. Using Pt as a reference, whose SHA is positive, we infer a positive effective SHA for RuO$_2$/Py and a negative one for RuO$_2$/YIG. For the RuO$_2$/Py, the maximum (minimum) voltage is observed at $\varphi_c=0^\circ$ ($\varphi_c=90^\circ$). The full angular dependence of the spin-pumping voltage is summarized in Fig.~\ref{fig1}(f). For both RuO$_2$/YIG and RuO$_2$/Py bilayers, the measured voltages consist of an isotropic offset superposed with an angular-dependent contribution. In both systems, the angular-dependent term follows a $\cos^2\varphi_c$ dependence, which is dictated by the crystal symmetry of the RuO$_2$(101) film and is consistent with previous reports~\cite{wang2024inverse,bose2022tilted,karube2022observation}. A pronounced contrast emerges in the isotropic offset, which exhibits opposite signs in the two bilayers. This sign reversal directly leads to positive and negative effective SHAs for RuO$_2$/Py and RuO$_2$/YIG, respectively, and unambiguously indicates competing spin–charge conversion mechanisms. The negative effective SHA in RuO$_2$/YIG is independently confirmed by spin-Seebeck measurement~\cite{SM}.
	
{\color{red}\it Identifying the physical mechanisms.---}Spin-charge conversion in RuO$_2$ stacks can originate from three distinct sources: the non-relativistic inverse ASSE, the relativistic ISHE in bulk RuO$_2$, and the interfacial IREE. Because the inverse ASSE requires crystalline order, it is absent in polycrystalline films and can therefore be excluded by using such samples. To disentangle bulk ISHE and interfacial IREE, we exploit their response to the direction of spin injection: the ISHE voltage reverses sign when the spin current is injected from the opposite side of the RuO$_2$ layer, whereas the IREE voltage remains unchanged~\cite{shen2021agbi,cheng2022coherent}. 

To enable simultaneous detection of spin currents from either interface {\it within a single measurement}, we fabricated a YIG/RuO$_2$/Py trilayer [Fig.~\ref{fig2}(a)], comprising epitaxial (111) YIG and polycrystalline Py, whose distinct FMR fields are well separated at 14~GHz. The 8-nm-thick polycrystalline RuO$_2$ layer was deposited at 450~$^\circ$C that possibly contained RuO$_2$(100) texture~\cite{SM}, and the inverse ASSE was definitely suppressed. Figure~\ref{fig2}(b) shows the measured voltage as a function of applied field for the YIG/RuO$_2$/Py trilayer. The voltage exhibits a double-peak profile at both positive and negative magnetic fields, corresponding to FMR of YIG ($\approx\pm4$~kOe) and Py ($\approx\pm2.2$~kOe). This double-peak profile reproduces the sign difference observed in Fig.~\ref{fig1}. The opposite signs of $\Delta V_{\rm SP}$ at YIG resonance in Fig.~\ref{fig2}(b) compared with Fig.~\ref{fig1}(d) arise from the inverted stacking order of RuO$_2$ and YIG~\cite{yu2020fingerprint}. For comparison, a YIG/Pt/Py control trilayer exhibits a characteristic peak-and-dip profile governed solely by the ISHE of Pt, as shown in Fig.~\ref{fig2}(c). Therefore, the double-peak structure in Fig.~\ref{fig2}(b) suggests a significant interfacial contribution in the YIG/RuO$_2$/Py system, because bulk ISHE alone would yield the same SHA sign for spin injection from either side.

\begin{figure}[b]
\centering
\includegraphics[width=\columnwidth]{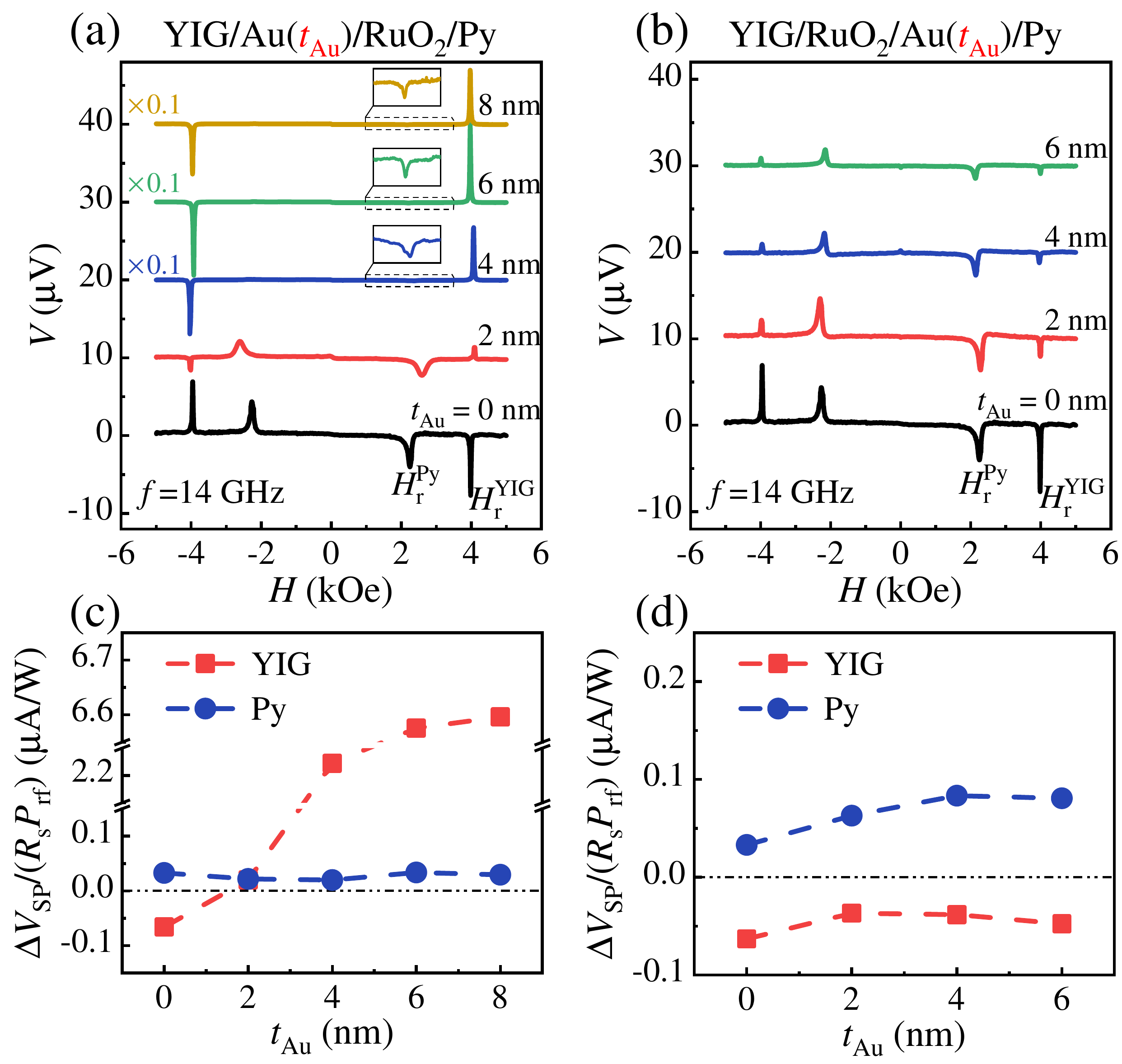}
\caption{Spin-pumping signals for YIG(42 nm)/RuO$_2$(8 nm)/Py(10 nm) with Au inserted at the (a) YIG/RuO$_2$ and (b) RuO$_2$/Py interfaces. (c, d) Normalized spin-pumping voltage $ \Delta V_{\mathrm{SP}}/(R_s P_{\mathrm{rf}})$ as a function of Au thickness, allowing quantitative comparison. \label{fig3}}
\end{figure}	
	
\begin{figure*}[htbp]
\centering
\includegraphics[width=14cm]{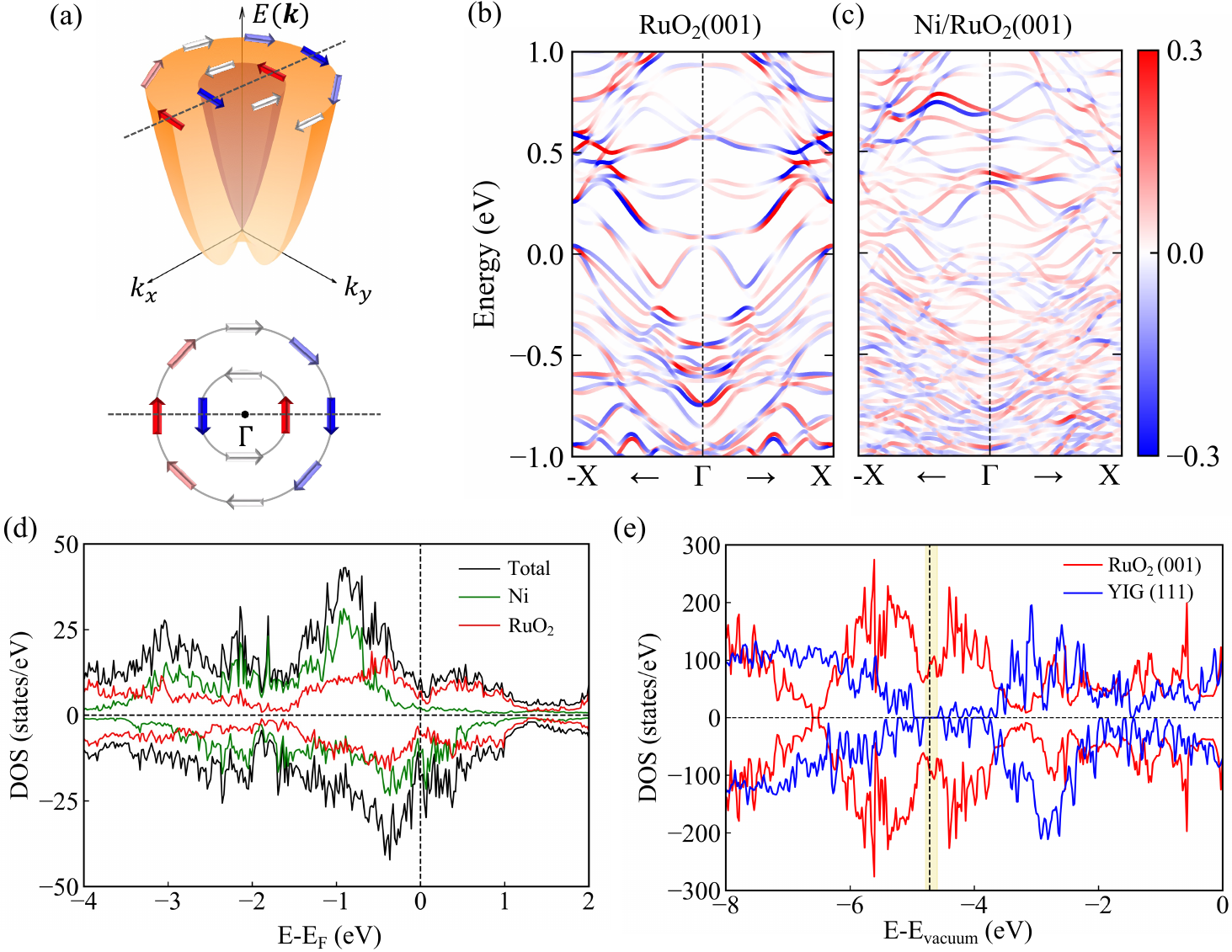}
\caption{(a) Schematic Rashba splitting and helical Fermi contours. Opposite momenta carry opposite transverse spins. (b) Spin-projected band structure of RuO$_2$(001). (c) Spin-projected band structure of RuO$_2$(001) with a Ni adsorption layer. (d) DOS of RuO$_2$(001) with Ni adsorption. SOC is turned off for clarity. (e) DOS of RuO$_2$(001) (red lines) and YIG(111) (blue lines), independently calculated with vacuum level aligned. The RuO$_2$ Fermi level (black dashed line) falls inside the energy gap of YIG (light yellow shadow), suppressing hybridization. The RuO$_2$ surface DOS is normalized to the same surface area as the YIG(111) unit cell. \label{fig4}}
\end{figure*}
		
To confirm the interfacial origin, we insert ultrathin Au layers at either RuO$_2$ interface. Au is ideal for this control experiment because it resists oxidation at high temperatures required for RuO$_2$ growth, possesses a negligible SHA~\cite{SM,qu2013intrinsic}, and maintains a long spin-diffusion length~\cite{isasa2015temperature,nair2021spin}, thereby acting as a spin-transparent spacer that suppresses IREE while transmitting the spin current. Inserting the Au layer between YIG and RuO$_2$ reverses the spin-pumping voltage at the YIG resonance [Fig.~\ref{fig3}(a)], indicating that the spin current from YIG is initially converted by the IREE at the YIG/RuO$_2$ interface. Once this interface is decoupled, the remaining signal arises from bulk ISHE in bulk RuO$_2$, whose positive SHA yields the opposite voltage. Conversely, the Py resonance signal retains its sign, confirming that spin current injected from Py is predominantly converted within the bulk before reaching the YIG/RuO$_2$ interface. Note that the resonance field of Py is slightly shifted with inserting Au, and this shift can be attributed to the microstructural changes in the stacks with the Au layer. Figure~\ref{fig3}(b) shows the voltage for YIG/RuO$_2$/Au/Py stacks as Au thickness at the RuO$_2$/Py interface is increased. No sign reversal is observed, corroborating that the RuO$_2$/Py interface contributes negligibly to spin-charge conversion. The positive SHA deduced from the Py side is therefore intrinsic to bulk RuO$_2$.

Peak heights in Fig.~\ref{fig3}(a) and (b) vary because the Au interlayer modifies microwave absorption. We therefore plot the normalized spin-pumping signal $ \Delta V_{\mathrm{SP}}/(R_s P_{\mathrm{rf}})$ as a function of Au thickness in Fig.~\ref{fig3}(c) and (d), where $R_s$ is the resistance of the stack and $P_{\mathrm{rf}}$ is the absorbed rf power~\cite{SM}. The normalized Py-resonance signal remains unchanged regardless of Au placement, underscoring bulk-dominated conversion. The YIG-resonance signal, however, flips sign when Au is introduced at the YIG/RuO$_2$ interface while remaining constant when Au is placed at the RuO$_2$/Py interface. This systematic evolution unambiguously reveals a competition between the IREE at the YIG/RuO$_2$ interface and the ISHE in the bulk RuO$_2$, with the former dominating in the absence of an Au spacer.

The monotonic increase of the red squares in Fig.~\ref{fig3}(c) reflects progressively diminished YIG/RuO$_2$ interfacial contact: ultrathin Au interlayers remain discontinuous owing to the high-temperature RuO$_2$ growth process, allowing residual IREE-mediated spin-charge conversion~\cite{SM}. In contrast, thicker Au films form uniform, continuous barriers that fully suppress the interfacial contribution, isolating the bulk ISHE from RuO$_2$. Independent thickness-dependent measurements confirm that the IREE dominates spin-charge conversion in RuO$_2$/YIG bilayers, whereas the bulk ISHE dominates in RuO$_2$/Py bilayers~\cite{SM}.

{\color{red}\it Rashba splitting of RuO$_2$ surface states.---}To elucidate the microscopic origin of the interface-selective spin-charge conversion, we performed first-principles calculations of the Rashba-type surface states of RuO$_2$. Figure~4(a) illustrates the canonical Rashba splitting induced by interfacial SOC with broken inversion symmetry, lifting spin degeneracy and producing concentric Fermi contours with opposite helicities~\cite{manchon2015new,soumyanarayanan2016emergent,bychkov1984properties}. In particular, the electronic states at both sides of $\Gamma$ point have opposite spin orientations perpendicular to the crystal momentum.
	
Figure~\ref{fig4}(b) presents the electronic bands of the RuO$_2$(001) surface near the Fermi level, with color denoting the in-plane spin component perpendicular to crystal momentum. The states display the expected antisymmetric spin texture around the $\Gamma$ point, confirming the presence of Rashba splitting and enabling efficient spin-charge conversion via the IREE~\cite{edelstein1990spin,inoue2003diffuse,miron2010current}. RuO$_2$(100) and (101) surfaces exhibit similar features~\cite{SM}. To model the RuO$_2$/Py interface, we adsorb two atomic layers of Ni on RuO$_2$(001) and the resulting band structure is shown in Fig.~\ref{fig4}(c). Metallic Ni introduces a dense manifold of states near the Fermi level and eliminates the characteristic antisymmetric spin texture around $\Gamma$, indicating strong hybridization that quenches the Rashba states. Projected density of states (DOS) calculations [Fig.~\ref{fig4}(d)] reveal comparable Ni and RuO$_2$ contributions at the Fermi level, corroborating the loss of spin-momentum locking and consistent with our experimental observation that the RuO$_2$/Py interface contributes negligibly to spin-charge conversion. Quenching of Rashba-type surface states by band hybridization is further substantiated by more realistic modeling of RuO$_2$/Py interface~\cite{SM}.

A full {\it ab initio} treatment of the RuO$_2$/YIG interface is very challenging due to the incommensurate lattice. Instead, we independently calculate the DOS of YIG(111) and RuO$_2$(001), as shown in Fig.~\ref{fig4}(e), aligning their vacuum level to zero. The Fermi level of RuO$_2$ surface (dashed line) lies approximately 5~eV below vacuum and falls within the energy gap of YIG surface (light yellow shadow). Consequently, RuO$_2$ surface states experience minimal hybridization with YIG, preserving their Rashba character and enabling a robust IREE.

Previous study reported topologically protected surface states in RuO$_2$~\cite{sun2017dirac} arising from bulk Dirac nodal lines (DNLs) along the [110] and [1$\bar{1}$0] directions of the Brillouin zone. These DNLs carry a nontrivial Berry phase of $\pi$, guaranteeing topological protection. However, SU(2) symmetry is broken once SOC is included, gapping the DNLs and lifting topological protection. Consequently, these surface states split into spin-polarized, Rashba-like bands. Lacking topological protection, these Rashba states are fragile and can be strongly quenched by electronic hybridization with adjacent metallic layers.

Our results indicate that Rashba states exist at the RuO$_2$ surface only when interfacial hybridization is weak. While bulk RuO$_2$ possesses a positive SHA, the significant IREE at the RuO$_2$/YIG interface contributes a large negative signal and inverts the sign of the spin-pumping voltage [red symbols in Fig.~\ref{fig1}(f)]. While spin-resolved angle-resolved photoemission spectroscopy has previously detected Rashba splitting in RuO$_2$~\cite{liu2024absence}, our transport measurements provide the first direct evidence of the resulting IREE. Notably, these conclusions remain unchanged regardless of whether RuO$_2$ hosts altermagnetic order or is nonmagnetic~\cite{SM}. 

{\color{red}\it Conclusions.---}Spin-pumping measurements reveal opposite effective SHAs in RuO$_2$/YIG and RuO$_2$/Py bilayers, irrespective of RuO$_2$ crystallinity. Inserting ultrathin Au at the RuO$_2$/YIG interface reverses the spin-pumping voltage, confirming that the negative SHA arises from a dominant IREE. In RuO$_2$/Py, by contrast, the conversion is governed by the bulk ISHE. First-principles calculations show that Rashba-split surface states survive at the RuO$_2$/YIG interface but are quenched by strong hybridization with metallic Py. These results provide a unified understanding of spin-charge conversion in the prototypical altermagnet RuO$_2$ and demonstrate that its efficiency and sign can be engineered simply by selecting the adjacent ferromagnet, offering a new design knob for low-power spintronic devices.

\acknowledgements
This work was supported by the National Key Research and Development Program of China (Grant No. 2022YFA1204002 and No. 2024YFA1408500) and the National Natural Science Foundation of China (Grants No. 12274323, No. 12574115, and No. 12374118).

\end{document}